\documentclass[1p,authoryear]{elsarticle}

\usepackage{amssymb}
\usepackage{amsmath}
\usepackage{graphicx}
\usepackage{natbib}
\usepackage{color}

\journal{Journal of Theoretical Biology}

\begin{document}

\begin{frontmatter}



\title{Giant number fluctuations in microbial ecologies}


\author[dip]{Dipjyoti Das \corref{cor1}}
\ead{dipjyoti@phy.iitb.ac.in}

\author[dip]{Dibyendu  Das}
\ead{dibyendu@phy.iitb.ac.in}

\author[ashp]{Ashok Prasad}
\ead{ashokp@engr.colostate.edu}

\cortext[cor1]{Corresponding author}

\address[dip]{ Department of Physics, Indian Institute of Technology Bombay,
              Powai, Mumbai - 400076, India}
%\address[dd]{ Department of Physics, Indian Institute of Technology Bombay,
             % Powai, Mumbai - 400076, India}

\address[ashp]{Department of Chemical and Biological Engineering,
        Colorado State University, Fort Collins, Colorado, USA.}

\begin{abstract}
Statistical fluctuations in population sizes of microbes may be quite
large depending on the nature of their underlying stochastic
dynamics. For example, the variance of the population size of a microbe undergoing a pure birth process with unlimited resources is proportional to the square of its mean. We refer to such large
fluctuations, with the variance growing as square of the mean, as Giant
Number Fluctuations (GNF).  Luria and Delbr\"uck showed that spontaneous mutation processes in microbial populations exhibit GNF. We explore whether GNF can
arise in other microbial ecologies. 
We study certain simple ecological models
evolving via stochastic processes:  (i) bi-directional mutation,
(ii) lysis-lysogeny of bacteria by bacteriophage, and
(iii) horizontal gene transfer (HGT). For the
case of bi-directional mutation process, we show analytically exactly that the
GNF relationship holds at large times.  For the ecological model of bacteria
undergoing lysis or lysogeny under viral infection, we show that if
the viral population can be experimentally manipulated to stay
quasi-stationary, the process of lysogeny maps essentially to  one-way
mutation process  and hence the GNF property of the lysogens follows.
Finally, we show that even the process of HGT may map to the mutation process at large times, and thereby exhibits GNF.
\end{abstract}


\end{frontmatter}

% \linenumbers

\section{Introduction}
\label{intro}

The presence of huge statistical fluctuations in microbial populations
was first shown in a simple biological process, namely the birth or
autocatalytic process that involves the binary fission of a cell into
two genetically identical daughters. The full mathematical solution of
the stochastic process with the exact distribution of the population
number, and in particular the variance, was calculated by 
\citet{delbruck40}. In such a process the mean of the population
number $n(t)$ grows exponentially with time $t$ as: $\langle n(t)
\rangle = n(0) \exp(\beta t)$, where $\beta$ is the birth rate, and
the variance $Var[n(t)]= \langle n(t) \rangle^2/n(0) - \langle n(t)
\rangle$, which grows as square of the mean at large times.  Due to
such huge fluctuations this process stands in direct contrast with
the familiar Poisson process which has $Variance = mean$. 
 In this paper, we refer to the statistically large fluctuations characterized by $ Variance \sim (mean)^{\nu}$ with $\nu=2$,  as `giant number fluctuations' (GNF). This term has been appeared previously in other contexts like spatial fluctuations of bacterial numbers \citep{swinney_EPL,swinney_PNAS}, where $\nu>1$ has been regarded as a signature of large fluctuations than usual. Here, since  we use the pure birth process as a reference case to compare with other ecological models, by GNF we specifically mean that the power $\nu=2$.
The GNF property continues to hold even for a more complicated
process, namely spontaneous forward mutation, as revealed by the
classic work of \citet{luria43}.

The work of \citet{luria43} (henceforth LD) in which a bacterial
population in a nutrient-rich medium was allowed to grow for a while,
and then subjected to an one-time attack of a lethal virus, was important
in many ways.  Their work provided evidence for spontaneous random
mutations in the DNA sequence of the genome \citep{molbio} and also
initiated a practical experimental approach to calculate the mutation
rate \citep{Rosche2000}. A key result of their work was the demonstration 
of GNF in the number $x(t)$ of the mutant bacteria at large
times --- $Var[x(t)] \sim \langle x(t) \rangle^2$. 
It is important to
note that GNF in the LD ecological model  does not trivially
follow from the results of a pure birth process. The mutant number in
an LD ecological model and the bacterial number in a pure birth process have
distinct distribution functions --- the former has a power law tail
\citep{Mandelbort,Ma,qizheng}, while the latter has an exponential
tail \citep{delbruck40}. Moreover the GNF property holds at large
times, and not at early times, in both the LD and the birth
processes. The LD model involves
three processes altogether, namely the birth process of the wildtype
bacteria, that of its mutant, and the mutation process.  \citet{luria43} treated the two birth processes  deterministically while the mutations were treated as random
events. \citet{lea-coul} improved the latter by also treating the birth of
mutants stochastically.  Finally Bartlett \citep[][pg. 23]
{Armitage,qizheng} obtained
the most realistic calculations by treating all the three processes as
stochastic. While all three formulations yield different distribution functions, the GNF property holds asymptotically for all of them. The LD result motivates us to explore the prevalence of the GNF property in other microbial
ecologies.
  
In a general microbial ecology, where complex inter-microbe
interactions are involved, one may ask: what is the relationship
between the variance and the mean of a particular species? In this
paper we study the asymptotic population dynamics of three progressively more
complex ecological models (as compared to the original LD case) namely (i) a model involving bi-directional mutation process, (ii) a particular limit of infection of
bacteria by bacteriophage and (iii) two simple models of horizontal gene transfer.

In the bi-directional mutation, bacteria not only mutate by a
forward mutation like the LD case, but the mutant can convert back to the original wildtype bacteria
\citep{revmut1}. Such reverse mutations are well known and in fact
form the basis of the Ames test which is one of the standard tests for
determining the mutagenicity of chemicals \citep{McCann1975,
  Mortelmans2000}. We show analytically exactly that the asymptotic
distribution of population sizes of both the wildtype and the mutant obeys the GNF relationship.

In nature, temperate bacteriophages (for example, phage $\lambda$ which infects  E. Coli) can
follow either one of two life cycles \citep{Sneppen,Sankar}. In the
lytic life cycle, the phage that infects the host cell multiplies
through replication and ultimately bursts out by killing the
bacterium. In the lysogenic life cycle the phage combines its genome
with that of the bacterium to form a stable lysogen \citep{kourilsky73,Weitz1,Goldenfeld10-phage-dynamics,Idogolding10}. Interestingly, we show that if experimentally the phage population size can be controlled to stay quasi-constant, this three-species system
reduces to an one-way mutation process involving two species. We
numerically demonstrate this LD like behavior and the associated GNF
property for our proposed protocol.

Finally we study  ecological models involving  horizontal gene
transfer (HGT). HGT occurs due to exchange of DNA segments of
a genome between different strains of bacteria, and can take place by
means of three known mechanisms, namely transformation, transduction,
or conjugation. The discovery of HGT as an important inter-microbe
interaction over the past few decades has led to a resurgence of
interest in microbial evolutionary dynamics \citep{hgtreview,hgt05,hgt08,goldWoese}. 
Here we have found that ecological models involving bacterial genotypes undergoing mutual HGT, may sometimes map to mutation processes at large times.  We numerically demonstrate the appearance of GNF in two- and four-species HGT models.

In all the ecological models we study in this paper, we assume a nutrient-rich
environment (like the experimental conditions of Luria and Delbr\"uck),
so that unrestricted birth is possible for each population for
indefinite time. Although in a natural environment the exponential
phase (often also called the ``logarithmic phase'') of the bacterial
growth may approach saturation due to overcrowding and limited food
resources, in a controlled laboratory experiment one can replenish
nutrients to prolong the exponential phase \citep[see][]
{revmut1}. The curious reader may wonder whether the GNF property
holds beyond the exponential growth phase or not. We will discuss
below the case of a birth process with ``limited resources'' and
deaths, and show that fluctuations are ordinary Poisson-like as
saturation is reached.

 Experiments that study microbial populations in the presence of
 unknown interactions among the different genotypes may therefore look
 for GNF. The prevalence of the GNF property in the systems that we 
 study, suggests, that it is likely to be found  often in the
 laboratory.

%%%%%%%%%%%%%%%%%%%%%%%%%%%%%%%%%%%%%%%%%%%%%%
\begin{figure*}
\centering
 \includegraphics[width=0.7\textwidth] {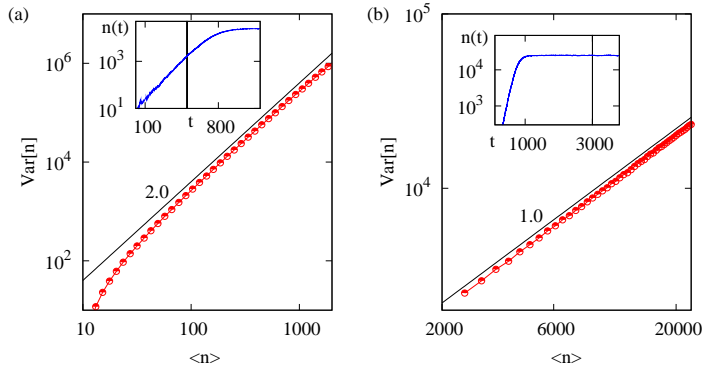}
\caption{The number fluctuations in the exponential phase and saturation phase 
of the logistic birth model are compared. (a) Log-log plot of $Var[n] \sim \langle n \rangle^2$ in the exponential phase, with $\langle n \rangle \sim \exp((\beta - \delta) t)$ (see inset), and (b) log-log plot $Var[n] \sim \langle n \rangle$ in the saturation phase, with $\langle n \rangle \sim  K$  (see inset). Note $\beta=0.02$, $\delta=0.01$, $n(0)=10$, and for the two insets $K = 50000$. Every data point (circles) of both (a) and (b) is obtained by averaging over $10^5$ independent histories. The data points in (a) are obtained at different chosen time $t$ for a fixed $K=50000$, while those in (b) are obtained at different chosen values of $K$ for a fixed time $t=3000$ which is deep inside the saturation (see inset). }  
\label{fig:0}       % Give a unique label
\end{figure*}
%%%%%%%%%%%%%%%%%%%%%%%%%%%%%%%%%%%%%%%%%%%%%%%%%%%%%

\section{Birth and bi-directional mutation processes}
\label{sec:1}

\subsection{Birth process}
\label{subsec:birth}

The process by which a bacterial species multiplies by successive
division is stochastic. In the presence of unlimited resources it is
represented as: $ N \stackrel{\beta}{\rightarrow} N + N$, where
$\beta$ is the birth rate. The probability $P_{n}(t)=Prob.[n(t) = n]$
of the random population size $n(t)$ at any time $t$ is governed by
the Master quation:
\begin{eqnarray}
\frac{\partial P_{n}(t)}{\partial t} &=& \beta (n-1)P_{n-1}(t) - \beta n P_{n}(t)
\label{birth:1}
\end{eqnarray}
The exact distribution $P_n(t)$ was solved by \citet{delbruck40}  and
that leads to: $Var[n] = \langle n \rangle^2 / n(0) - \langle n
\rangle$. At large times, $Var[n] \sim \langle n \rangle^2$, which is
the GNF property.

In the presence of limited resources the bacterial population size is
expected to saturate. Does the GNF property hold in the saturation
phase?  A standard model of microbial growth under nutrient-limiting conditions is the logistic growth model, where population numbers eventually saturate due to limited resources. We studied a stochastic version of the logistic growth model \citep{logis_birth1} represented by the following processes:
\begin{eqnarray}
R_1: &&~~ N \stackrel{\beta(1-n/K)}{\xrightarrow{\hspace*{0.8cm}}} N + N \nonumber \\
R_2: &&~~ N \stackrel{\delta}{\longrightarrow} 0 
\label{birth:2}
\end{eqnarray}
Here, the parameter $K$ is called the `carrying capacity' and represents the maximal population that can be supported by the available resources. Note that
the birth rate $\beta^{'}\equiv \beta(1-n/K) $ is not constant. The
inclusion of death (reaction $R_2$ in Eq. (\ref{birth:2})) is
important --- without it the evolution in the saturation phase would
stall as $n(t) \rightarrow K$ (and $\beta^{'} \rightarrow 0$). The Master equation of this process is non-linear in population size:
\begin{eqnarray}
\frac{\partial P_{n}(t)}{\partial t} &=& \beta(1-(n-1)/K) (n-1)P_{n-1}(t)+\delta(n+1)P_{n+1}(t)\nonumber\\
&& - (\beta(1-n/K)+\delta)n P_{n}(t)
\label{birth:3}
\end{eqnarray}

We simulate Eq. \ref{birth:2} using kinetic Monte-Carlo
\citep{Kalos75,Gillespie76,Gillespie07} and the results for $Var[n] =
\langle n^2 \rangle - \langle n \rangle^2$ in the exponential and
saturation phases of the system, are shown in Fig. 1(a) and
 1(b), respectively. While (a) shows GNF, (b) shows $Var[n] \sim
\langle n \rangle$. In this model, a small proportion of the stochastic histories lead to extinctions of the population ( $\sim 0.1 \%$ for $K=5000$ for example).
 We obtain $Var[n]$ by doing
a history average over cultures which do not become extinct, starting
from a fixed $n(0)$. The result in Fig. 1(b) shows that the giant fluctuations 
which arise in the exponential growth phase, diminish and become
ordinary Poisson-like in the saturation regime.

%%%%%%%%%%%%%%%%%%%%%%%%%%%%%%%%%%%%%%%%%%%%%%%%%%%%%%%%%%%%%%%%%%%%%%%%%%%%

\begin{figure*}
\centering
  \includegraphics[width=0.65\textwidth]{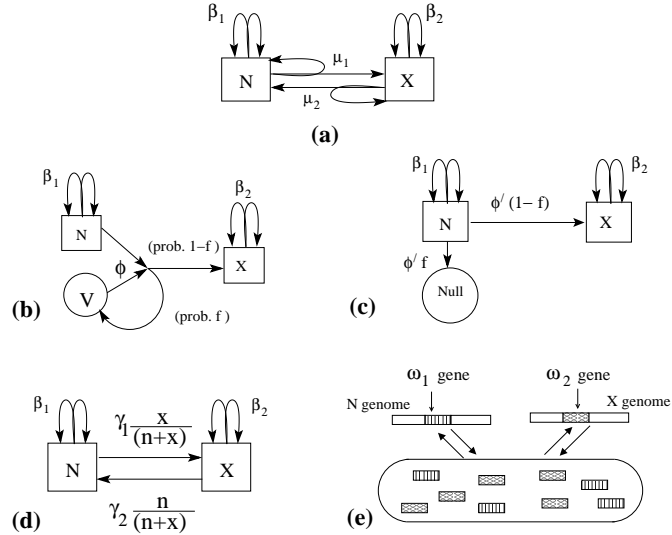}
\caption{The schematic diagrams of the  ecological models. The double arrows going up from a population (marked by $\beta_1$ or $\beta_2$) represent the birth processes.  (a) Bi-directional mutation. (b) Lysis-Lysogeny.  (c) The LD limit of the Lysis-Lysogeny model with
  constant viral population. (d) Two species HGT. (e) Visualization of
  the two species HGT mechanism via a shared gene pool. }
\label{fig:1}       % Give a unique label
\end{figure*}
%%%%%%%%%%%%%%%%%%%%%%%%%%%%%%%%%%%%%%%%%%%%%%%%%%%%%%%%%%%%%%%%%%%%%%%%%%%%%%

\subsection{Bi-directional mutation}
\label{subsec:mut}

In this section we calculate the moments of the population sizes of
species involved in a bi-directional mutation process. We follow the
method of Bertlett \citep[pg. 34--40][pg. 23]
{Armitage,Bertlett,qizheng} in which the mutation process is assumed
to produce a mutant and a normal cell from a normal cell (formulation
(B) of D. G. Kendall,  \citep[see][pg. 4]{qizheng}).  In nature there
exists a possibility of an occasional reverse mutation that takes the
microbe to the original genotype \citep{revmut1}, thereby diminishing
the population of the mutant pool.  The studies on spontaneous forward
and reverse mutation of different bacterial genotypes have revealed
that the latter two rates may differ by a factor of $10 - 100$ in
magnitude \citep{revmut2,revmut1}. Let the wildtype population be
denoted by $N$ (with a population size $n(t)$). The mutant population $X$ 
(with a population size $x(t)$) can convert back to the
wildtype population $N$ (see Fig. 2a)  via the reaction channel
$R_4$ in Eq. (\ref{revmut:1}), which makes the problem two-way
coupled. The reactions are: 
 \begin{eqnarray}
R_1: &&~~ N \stackrel{\beta_1}{\rightarrow} N + N \nonumber \\
R_2: &&~~ X \stackrel{\beta_2}{\rightarrow} X + X \nonumber \\
R_3: &&~~ N \stackrel{\mu_1}{\rightarrow} N + X \nonumber \\
R_4: &&~~ X \stackrel{\mu_2}{\rightarrow} N + X .
\label{revmut:1}
\end{eqnarray}
Here $\beta_1$ and $\beta_2$ are the bare birth rates of the wildtype
and mutant population respectively, and $\mu_1$ and $\mu_2$ are the
forward and the reverse mutation rates respectively.

We begin by considering the joint probability distribution
$P_{n,x}(t)=Prob.[n(t) = n ; x(t) = x]$. From the rates given in
Eq. (\ref{revmut:1}), the master equation is as follows:

\begin{eqnarray}
\frac{\partial P_{n,x}(t)}{\partial t} &=& \beta_1 (n-1)P_{n-1,x}(t) +
\beta_2(x-1) P_{n,x-1}(t) +\mu_1 n P_{n,x-1}(t)+\mu_2 x
P_{n-1,x}(t)\nonumber\\ && -(\beta_1 n +\beta_2 x +\mu_1 n +\mu_2
x)P_{n,x}(t)
\label{revmut:2}
\end{eqnarray}

The above equation can be solved analytically exactly for its moments
using the generating function technique (see \ref{sec:App_mut}).
 We find that the means of both $N$ and $X$
populations follow the same time development at asymptotically large
 times (see Eq. \ref{revmut:moments} in Appendix A):
\begin{equation} 
\langle n \rangle \sim \langle x \rangle \sim e^{\alpha_1 t}, ~~\text{where,}~\alpha_1 = [ (\beta_1 + \beta_2) +\sqrt{(\beta_1 - \beta_2)^2 + 4 \mu_1 \mu_2}]/2
\label{revmut:mean} 
\end{equation}
Thus the bi-directional coupling makes the wildtype and mutant track
each other in perfect synchrony. The effective growth rate $\alpha_1$
of the two species is a complicated combination of all the rates defining
the ecological model.
Turning to the fluctuations (see Eq. \ref{revmut:moments} in Appendix A), we find that mathematically the largest exponent contributing to the variance is such that at large $t$:
\begin{equation} 
 Var[n] \sim Var[x]\sim e^{2 \alpha_1 t}.
\label{revmut:var} 
\end{equation}
 Thus  the GNF property holds: $Var[n] \sim
\langle n \rangle^2$ and $Var[x] \sim \langle x \rangle^2$. We note that this GNF property holds as long as there is no saturation.

\section{LD limit of Lysis-Lysogeny}
\label{sec:lysis}

The prolonged co-evolution of  temperate phage with a bacterial population can be modeled as a stochastic process  wherein virus-infected  bacteria either  die by lysis, or transform into another genotype via lysogeny \citep{Sneppen,Sankar}. A large number of virus particles are born during lysis.  The  classic host-phage system of the bacteria E. Coli and phage $\lambda$ exhibits the processes of lysis and lysogeny \citep{Idogolding10}. This ecology can be modeled  by the following reactions (see Fig. 2b):

\begin{eqnarray}
R_1:&& ~~ N \stackrel{\beta_1}{\rightarrow} N + N \nonumber \\
R_2:&& ~~ N + V  \stackrel{\phi f}{\longrightarrow}\alpha V  \nonumber \\
R_3:&& ~~ N + V  \stackrel{\phi (1-f)}{\xrightarrow{\hspace*{0.8cm}}}X  \nonumber \\    
R_4:&& ~~ X \stackrel{\beta_2}{\rightarrow} X + X 
\label{lysis:1} 
\end{eqnarray}

Here $N$, $V$ and $X$ denote the sensitive bacteria, the virus
particles, and the lysogenic bacteria that is immune to the virus respectively, and
$n(t)$, $v(t)$, and $x(t)$ are their respective population sizes.  The
sensitive bacteria grows at rate $\beta_1$. The viral infection rate,
i.e. the rate of a $V$ interacting with an $N$, is $\phi$.  After
infection, the probability of lysis is $f$ and that of lysogeny is
$(1-f)$.  On lysis, the bacterium dies and the number of viruses
released is $\alpha$ (the average viral burst size).  On lysogeny, 
a stable new lysogen is formed which grows at a distinct rate
$\beta_2$. We assume that only one virus infects a bacterium at a
time, i.e. the multiplicity of infection (MOI) is one. Recent
experiments have shown that the probability of lysogeny is $\approx 0.2$
when the MOI equals one \citep{Idogolding10}, though older experiments
had indicated it to be negligible at this MOI \citep{kourilsky73}.
One can write the Master equation for the joint probability
$P_{n,x,v}(t)=Prob.[n(t) = n;~x(t) = x;~v(t)=v]$ describing the
ecological model in Eq.(\ref{lysis:1}) (see Eq.(\ref{appB:1}) in Appendix B);
but because of it having {\it nonlinear} terms, unlike
Eqs. (\ref{birth:1}) and (\ref{revmut:2}) in Sec. (\ref{sec:1}), it is
hard to tackle analytically. We note that in the above we are ignoring the possibilities of delay of infection to lysis, prophage induction and curing of lysogens, and superinfection.

It is well known \citep{Weitz-phage-dynamics,Goldenfeld10-phage-dynamics} that without
a finite carrying capacity of bacteria due to limited resources, a model like in Eq. (\ref{lysis:1}) makes all the sensitive bacteria
$N$ go extinct. The reason is that $\alpha$ is typically large (say
$\gtrsim 50$), and in a well mixed medium, the viral population
increases very fast, and infects and kills all the sensitive bacteria,
leaving the lysogens alive which continue to replicate on their
own. Apart from finite carrying capacity, another way to sustain this
model ecology is to make the viral population size $v(t)$ constant by some
means. This cannot be done in an exact sense, because that would
require viral bursts to stop.  But this may be done in an approximate way by a plausible experimental protocol that involves replacing part of the culture media by fresh media (see Appendix C).

If the viral population is kept fixed, under the condition
$v(t)=constant$, the model defined by Eq. (\ref{lysis:1}) reduces to
a two-species model (see Fig. 2c) represented by
\begin{eqnarray}
R_1:&& ~~ N \stackrel{\beta_1}{\rightarrow} N + N \nonumber \\
R_2:&& ~~ N \stackrel{\phi^{'}f}{\longrightarrow} 0 \nonumber \\
R_3:&& ~~ N\stackrel{\phi^{'}(1-f)}{\xrightarrow{\hspace*{0.8cm}}} X \nonumber \\
R_4:&& ~~ X \stackrel{\beta_2}{\rightarrow} X + X  ,
\label{lysis:2}
\end{eqnarray}
where, $\phi^{'}=\phi v = {\rm constant}$ is a new rate.  The above
reactions represent an LD-like ecological model with a
Master equation given by Eq.(\ref{appB:2}).  The reaction $R_3$ in
Eq. (\ref{lysis:2}) is like an uni-directional mutation reaction with
a mutation rate $\mu^{'} = \phi^{'}(1-f)$. Yet the mutation process $N
\rightarrow X$ is distinct from the process $N \rightarrow N+X$ used
in Berlett's formulation of mutation \citep[][pg. 23]
{Bertlett,qizheng} --- they have been referred to as formulations (A)
and (B) of D. G. Kendall \citep[see][pg. 4]{qizheng} respectively. Apart from this distinction
from the LD ecological model, another point of difference is the death of the
bacteria $N$ (reaction $R_2$). The latter leads to an effective growth
rate $\beta_1^{'} \equiv (\beta_1 - \phi^{'})$ of $N$. We present the
explicit results of the moments of population sizes for Eq.
(\ref{lysis:2}) in Eqs. (\ref{appB:3}) and (\ref{appB:4}).  The
results differ from those of the original LD problem only upto the constant
prefactors, which do not affect the time dependences. The means and
the variances are like LD at large times. Thus for
$\beta_1^{'}>\beta_2$, $\langle n \rangle \sim \langle x \rangle \sim
e^{\beta_1^{'} t}$ and $Var[n]\sim Var[x] \sim e^{2 \beta_1^{'}t}$,
while for $\beta_2 > \beta_1^{'}$, $\langle n \rangle \sim
e^{\beta_1^{'}t}$, $\langle x \rangle \sim e^{\beta_2 t}$ and $Var[n]
\sim e^{2 \beta_1^{'}t}$, $Var[x] \sim e^{2 \beta_2 t}$. Thus in all
cases, the GNF property holds at large times, i.e. $Var[n] \sim
\langle n \rangle^2$ and $Var[x] \sim \langle x \rangle^2$, for the
lysis-lysogeny model if $v(t)$ is held constant.

We suggest a way to experimentally hold the viral population $v(t)$ quasi-stationary (see Appendix C) by ``diluting out'' the viruses  \citep{kourilsky73} after equal time intervals $T$.
We implement the latter protocol numerically by resetting the viral population number  to its initial value $v_0$ after each interval $T$. In the meantime over $T$, we time-evolve the population numbers according to the original Lysis-Lysogeny reactions in Eq. (\ref{lysis:1}) using the
kinetic Monte-Carlo algorithm \citep{Kalos75,Gillespie76,Gillespie07}. Thus over a coarse time scale that is greater than $T$, the viral population $v(t)$ is quasi-stationary. We thus find  a time window within which this ecological model follows the LD limit.

\begin{figure*}
\centering
 \includegraphics[width=0.6\textwidth] {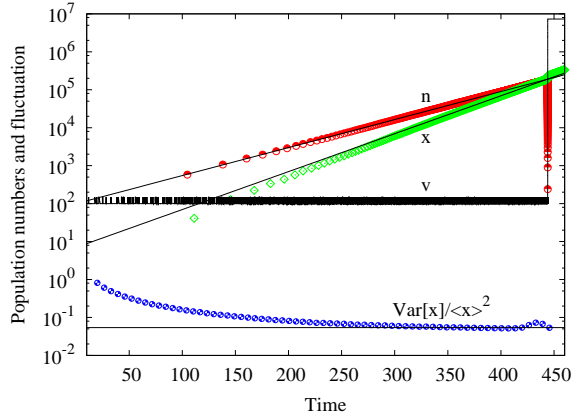}
\caption{Linear-log plot of $n(t)$, $x(t)$ and $v(t)$ versus time $t$
  for a single stochastic history. The fitted exponential functions
  (solid continuous straight lines) are: $\langle n \rangle \sim 100
  \times \exp(\beta_1^{'} t)$ (see section 3) and $\langle x \rangle
  \sim 7 \times \exp(\beta_2 t)$.  At the bottom we have
  $Var[x]/\langle x \rangle^2$ which is essentially constant between
  $t=200 - 400$, demonstrating the GNF property --- for  $Var[x]$ and $\langle x \rangle$,  averages over $10^5$ independent histories were done.  The data are for
  $\beta_1=0.02$, $\beta_2=0.023$, $\phi=0.00003$, $f=0.8$,
  $\alpha=50$, $T=0.001$, $n(0)=v(0)=100$ and $x(0)=0$. }
\label{fig:2}       % Give a unique label
\end{figure*}

We see in Fig. 3 that the data for the instantaneous
population sizes $n(t)$ and $x(t)$ of the bacteria and the lysogens,
match reasonably well with the growth curves of the means ($\langle n
\rangle$ and $\langle x \rangle$) predicted for the LD like problem of
Eq. \ref{lysis:2}.  In particular, the fluctuations of the lysogens
follow the GNF property as shown in Fig. 3 for large times
(within the window $t=200-400$), although not at small times.  The
time window over which the system behaves like LD may be made larger by
choosing a smaller interval $T$. Note that the window is eventually
terminated by a sharp catastrophic rise in the viral number as can be
seen around $t \sim 450$ in Fig. 3 (for a discussion see Appendix C).

In summary, we have proposed an  experimental protocol
which can make a stochastically evolving lysis-lysogeny ecology appear
as an LD ecology, over a certain time window. As a result, the GNF
property of the LD model  carries over to this special case of
lysis-lysogeny.  This GNF property holds only when there is no finite carrying capacity for the populations.

\section{Horizontal gene transfer}
\label{sec:HGT}

It is known that HGT produces a high degree of genomic similarity in
the interacting microbes \citep{hgtreview,goldWoese}. In the spirit of
a model studied recently for HGT for four genotypes within a biofilm
\citep{Goldenfeld08} we have constructed a simpler model of HGT
involving two bacterial genotypes, $N$ and $X$, with birth rates
$\beta_1$ and $\beta_2$ respectively. Population sizes are denoted by
$n(t)$ and $x(t)$. It is assumed that $N$ and $X$ are closely related
and they differ at the level of only one gene --- $N$ has a gene
$\omega_1$ and $X$ has a different gene $\omega_2$, while the rest of
the genome is identical for both. One may think of $N$ and $X$ as the
populations of a normal bacterium and an antibiotic resistant
variant. The antibiotic resistant bacterium has a mutation in one gene
and it can share the mutated gene by HGT and thereby confer antibiotic
resistance to the normal bacterium
(\citet{Akiba1960,Barlow2009}). Similarly the normal bacterium can
also share its copy of the gene and make the antibiotic resistant
bacterium revert back to the normal state. The genomic matters of $N$
and $X$ are exchanged via HGT, transforming one genotype completely
into other. This ecological model (see Fig. 2d) is described by the
following reactions :
\begin{eqnarray}
R_1: &&~~ N \stackrel{\beta_1}{\rightarrow} N + N \nonumber \\
R_2: &&~~ X \stackrel{\beta_2}{\rightarrow} X + X \nonumber \\
R_3: &&~~ N \stackrel{\gamma_1 x/(n+x)}{\xrightarrow{\hspace*{0.8cm}}} X \nonumber \\
R_4: &&~~ X \stackrel{\gamma_2 n/(n+x) }{\xrightarrow{\hspace*{0.8cm}}} N
\label{HGT:1}
\end{eqnarray}

In the above, the processes of HGT are viewed as two-step processes.
The genes are exchanged via a communally shared gene-pool (see
Fig. 2e). Firstly, a particular species can get a suitable
gene from the shared gene pool with a certain probability, which is
assumed to be equal to the fraction of its contributor. Thus a gene $\omega_2$ contibuted by $X$ is acquired with
probability $x/(n+x)$, and $\omega_1$ contributed by $N$ is acquired
with probability $n/(n+x)$.  Secondly, the newly acquired gene is
pasted (or overwritten) upon the former one with rates $\gamma_1$ and
$\gamma_2$.  Thus the $N$ genome gets transformed into a $X$ genome at
an effective rate of $\gamma_1 x/(n+x)$ ($R_3$ in Eq. (\ref{HGT:1})),
while $ \gamma_2 n/(n+x)$ is the effective rate at which $X$ genome
gets transformed into a $N$ genome ($R_4$ in Eq. (\ref{HGT:1})); these
effective rates are no longer constants like in the LD model and signify
that the processes are nonlinear.  Since, HGT is a relatively rare event
in reality, we further assume
$\{\beta_1,\beta_2\}>>\{\gamma_1,\gamma_2\}$ in the subsequent
discussion.

One can write the Master equation for the joint probability $P_{n,x}(t)=Prob.[n(t) = n;~ x(t) = x]$ for the HGT model in Eq. \ref{HGT:1} as below:
\begin{eqnarray}
\frac{\partial P_{n,x}(t)}{\partial t} &=& \beta_1 (n-1)P_{n-1,x}(t) + \beta_2(x-1) P_{n,x-1}(t) \nonumber\\
&&+\gamma_1 [(n+1)(x-1)/(n+x)] P_{n+1,x-1}(t) \nonumber\\
&&+\gamma_2 [(n-1)(x+1)/(n+x)]  P_{n-1,x+1}(t)\nonumber\\
&&-(\beta_1 n +\beta_2 x +(\gamma_1 +\gamma_2) nx/(n+x))P_{n,x}(t)
\label{HGT:2}
\end{eqnarray}
Note that the above Master equation has nonlinear terms unlike
Eqs. (\ref{birth:1}) and (\ref{revmut:2}) in Sec. \ref{sec:1} and
is difficult to tackle analytically.  One cannot write a system of ODEs
for the cumulants in a closed form, and in fact, following similar steps as in
Appendix A, every ODE for a cumulant contains the next higher order
cumulant. In spite of this difficulty, we will show that the unique
algebraic form of non-linearities in the HGT model makes its
asymptotic large time limit tractable.  At large times, both the
populations ($N$ and $X$) are very large and one of them dominates
over the other.

\begin{figure*}
\centering
  \includegraphics[width=0.7\textwidth]{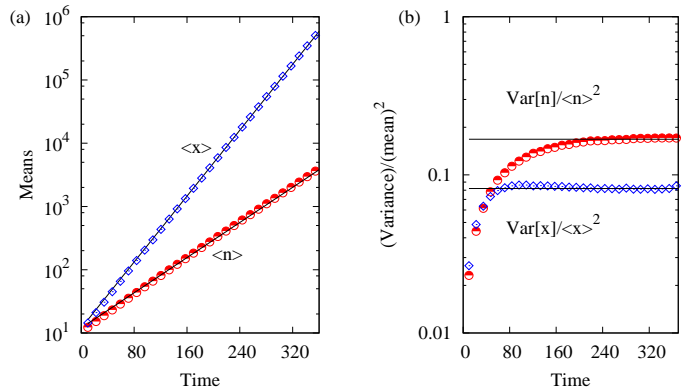}
\caption{ (a) Linear-log plot of $\langle n \rangle$ and $\langle x
  \rangle$ versus time $t$ for two-species HGT. The data from kinetic
  Monte-Carlo simulation are shown in symbols. The fitted exponential
  functions (solid continuous straight lines) are: $\langle n \rangle
  \sim 12.0 \times \exp((\beta_1-\gamma_1+\gamma_2)t)$ and $\langle x
  \rangle \sim 12.0 \times \exp(\beta_2 t)$ (see Sec. \ref{sec:HGT}).
  (b) Linear-log plot of $Var[n]/\langle n \rangle^2$ and $Var[x]
  /\langle x \rangle^2$ versus time $t$ for two-species HGT. The
  straight lines show that the data is constant at large times. The
  data in both the curves are for $n(0)=x(0)=10$, $\beta_1=0.02$,
  $\beta_2=0.03$, $\gamma_1=0.005$ and $\gamma_2=0.001$. All data are obtained by averaging over $10^5$ independent histories.}
\label{fig:3}       % Give a unique label
\end{figure*}
%%%%%%%%%%%%%%%%%%%%%%%%%%%%%%%%%%%%%%%%%%%%%%%%%%%%%%%%%%

We will discuss one case in detail, namely when $X$ dominates over
$N$ (i.e. $x>>n$ in Eq. (\ref{HGT:2})), which occurs for $\beta_2 >
\beta_1$ at asymptotically large times.  Under this condition, $(x\pm
1)/(n+x) \approx x/(n+x) \approx 1$ in Eq.  (\ref{HGT:2}), and one can
get an {\it approximate} linear Master equation:
\begin{eqnarray}
\frac{\partial P_{n,x}(t)}{\partial t} &\approx& \beta_1
(n-1)P_{n-1,x}(t) + \beta_2(x-1) P_{n,x-1}(t) +\gamma_1 (n+1)
P_{n+1,x-1}(t) \nonumber\\ 
&&+\gamma_2(n-1) P_{n-1,x+1}(t) -(\beta_1 n
+\beta_2 x +(\gamma_1 +\gamma_2) n)P_{n,x}(t)
\label{HGT:3}
\end{eqnarray}
Note that the above Eq. (\ref{HGT:3}) is similar to the Master equation of the
LD like ecological model discussed in Appendix B (see Eq. (\ref{appB:2})) ---
they are not exactly identical as the third and fourth terms in the two
equations are different. Starting from Eq. (\ref{HGT:3}) we have
derived the ODEs of the cumulants and compared them with the
Eq. (\ref{appB:3}) in Appendix B --- they are very similar, but not
identical. After making the identifications: $\gamma_1-\gamma_2 \equiv
\mu^{'}$, $\beta_1 - (\gamma_1-\gamma_2)\equiv \beta_1^{'}$ and
$\gamma_1+\gamma_2\equiv\phi^{'}$, we find that the new ODEs of
$\kappa_{1,0}$, $\kappa_{0,1}$ and $\kappa_{2,0}$ in the asymptotic HGT
model are same as those of Eq. (\ref{appB:3}), but ODEs of
$\kappa_{1,1}$ and $\kappa_{0,2}$ have minor differences. The latter
differences do not change the asymptotic temporal behavior. Solving the
ODEs, at large times we get: $\langle n \rangle \sim e^{
  (\beta_1-(\gamma_1-\gamma_2))t}$, $\langle x \rangle \sim e^{\beta_2
  t}$ and $Var[ n ]\sim e^{ 2(\beta_1-(\gamma_1-\gamma_2))t}$, $Var[x]
\sim e^{2 \beta_2 t}$ for the parameter regime $\beta_2>\beta_1>>
\gamma_1 >\gamma_2$. Results for other parameter regimes, as well as
the opposite case of $n(t)>>x(t)$ will not be discussed, as they are
easy to derive following similar steps as above.

Instead of an approximate analysis for large times as above, one may
simulate the original exact non-linear model of the HGT process
(Eq. (\ref{HGT:1})) using kinetic Monte-Carlo for all times. We show
our data in Fig. 4 --- we find a  match with the
LD like behavior (as predicted from our approximate analysis above)
for the means (Fig. 4a). At short times  GNF is not seen (Fig. 4b), although the means grow exponentially at those times (see Fig. 4a).  As time gets
larger, the variances  exhibit the GNF property
(Fig. 4b) for both $N$ and $X$ populations.  Thus we arrive
at a conclusion --- the two species HGT process
approximates a mutation process at large times, due to the algebric
uniqueness of its non-linear interaction and has giant fluctuations in
its population sizes. Does this conclusion extend to the HGT process
involving more than two species?
 To answer this, we investigate a HGT model involving four genotypes studied by \citet{Goldenfeld08} and confirm that the GNF property holds in this model (see the detailed discussion in Appendix D).

For multiple, i.e. $s$ (with $s \geq 2$) number of species undergoing
HGT, the Master equation in general will involve non-linear terms
proportional to $(\sum_{i=1}^{r}a_i n_i/\sum_{i=1}^{s}b_i n_i)\times
n_j$, where $r<s$, $n_i$ is the population size of the $i^{th}$
species, and $\{a_i,b_i\}$ are constants. At large times such
non-linear terms may reduce to approximate linear forms for a certain
suitable choice of the parameters. This linearization would map the
original non-linear processes of HGT to linear mutation processes.  As
a consequence, an LD like behavior with giant fluctuations may show up
in certain parameter regimes of the HGT model with multiple species.

%\begin{figure*}
%\centering
% \includegraphics[width=0.7\textwidth]{4speciesHGT.eps}
%\caption{(a) Linear-log plot of $\langle n_{00} \rangle$, $\langle
 % n_{01} \rangle$, $\langle n_{10} \rangle$ and $\langle n_{11}
 % \rangle$ versus time $t$ for four-species HGT. Data from kinetic
%  Monte-Carlo simulation are shown in various symbols as
%  indicated. The fitted exponential functions (solid continuous
%  straight lines) are: $\langle n_{00} \rangle \sim 10.0 \times
%  \exp(\lambda_1 t)$; $\langle n_{01}\rangle \sim 10.5 \times \exp(\lambda_2
%  t)$; $\langle n_{10}\rangle \sim 8.6 \times \exp(\lambda_2 t)$ and
%  $\langle n_{11}\rangle \sim 12.0 \times \exp(\lambda_4 t)$ (see 
%  Eq. (\ref{Goldenfeld:3}) in Appendix C). (b)
%  Linear-log plot of $Var[n_{00}]/\langle n_{00} \rangle^2$,
%  $Var[n_{01}]/\langle n_{01} \rangle^2$, $Var[n_{10}]/\langle n_{10}
%  \rangle^2$ and $Var[n_{11}]/\langle n_{11} \rangle^2$ versus time
%  $t$ for four-species HGT. The straight lines show that the data are constant 
%   with time. The  data are for $n_{00}(0)=n_{01}(0)=
%  n_{10}(0)=n_{11}(0)= 10$, $\beta_1=0.03$, $\beta_2=\beta_3=0.02$,
%  $\beta_4=0.01$, $r=0.0005$, $q_0=0.9$ and $q_1=0.1$. The data are obtained by% averaging over $10^5$ independent histories. }
%\label{fig:4}  % Give a unique label
%\end{figure*}
%%%%%%%%%%%%%%%%%%%%%%%%%%%%%%%%%%%%%%%%%%%%%%%%%%%%%%%%%%%%

\section{Discussion}

In this paper, we have studied the property of giant number
fluctuation (GNF) in a few interesting microbial ecological models. While this is well-known for the models of the birth process and of uni-directional
mutation, we have shown that it also holds for bi-directional
mutation, and may  be seen under suitable conditions in ecologies
of microbes evolving via HGT and Lysis-Lysogeny. 
 In all the  ecological models that we have studied, the GNF property appears to arise because the populations effectively obey linear Master equations in population sizes.  Yet it is interesting to  identify the regimes in these interacting models where the stochastic dynamics resembles the ``LD-like'' mutation process. 

 We have shown that whenever
an ecological model effectively reduces to a LD-like ecological model at the level of the Master equation, the GNF property follows. In the Lysis-Lysogeny
model, given  the non-linear terms in the Master equation, making a prediction about GNF becomes difficult. Yet
in the special limit when the viral population may be controlled to be
quasi-stationary, we see that this process has fluctuation properties
 like a mutation process. Similarly for the HGT model, the
non-linear terms of the Master equation sometimes (and for two-species
HGT always) are of an algebraic form that at late times approach
linear limiting forms. This again implies an LD like stochastic
dynamics with an associated GNF property. The results for
the saturation phase of the stochastic logistic birth process,
suggest that exponential growth is  necessary for seeing GNF.  Whether a linearized Master equation involving multiple species would always lead to GNF, remains an open mathematical question to explore in future.

Recently, a similar large fluctuation property  has been observed in  {\it active  matter} systems \citep{swinney_EPL,swinney_PNAS}, where the non-equilibrium
dynamics of swarming bacteria produces local spatial number
fluctuations within a fixed volume, which are non-Poissonian.
We would like to comment that in such experiments \citep{swinney_PNAS}, if the
measurements are done over time periods greater than the division time
scale (say $\sim 30$ min) of the bacteria, the sources of large fluctuations would be twofold. The first source of this large fluctuation  would be non-equilibrium collective dynamics, while the second would be the stochastic
microbial population growth itself (as pointed out in
this paper). Thus the current work may be of relevance for
quantitative studies of microbial systems executing active motion over
long times. Apart from that, the GNF in microbial ecologies is not
just a mathematical curiosity, but has practical significance.  What
constitutes an asymptotic regime in practice depends upon the system
--- it is achieved when the time $t$ is much larger than the inverse
of the largest growth rate in the system. For bacteria with a high
dividing rate, this may be reached in only a few hours or a day of
culture, and the GNF therefore is of practical experimental
importance. 
Luria and Delbr\"uck's work gave rise to fluctuation assays, where mutation rates are estimated by experimentally determining the distribution of mutants in multiple parallel cultures \citep{Rosche2000}. As pointed out in \citet{Rosche2000}, alternate methods based on estimation of the mutant fraction in bacterial cultures are very unreliable due to the presence of GNF.  Measurements of quantities that are related to total population numbers of cells in exponential growth are susceptible to such large variances, and more careful analysis of fluctuations may be warranted. Since we have now shown that GNF may arise in other interesting ecological models, more careful analysis of the population sizes in these ecologies need to be done by theorists
and experimentalists in future.

%%%%%%%%%%%%%%%%%%%%%%%%%%%%%%%%%%%%%%%%%%%%%%%%%%%%%%%%%
\section*{Acknowledgements}
Dibyendu Das thanks American Physical Society for a travel grant to visit CSU (in $2011$), where part of this paper was written. Dipjyoti Das would  like to acknowledge Council of Scientific and Industrial Research (CSIR), India (JRF Award No. - 09/087(0572)/2009-EMR-I) for the financial support.

%%%%%%%%%%%%%%%%%%%%%%%%%%%%%%%%%%%%%%%%%%%%%%%%%%%%%%%%%%%%%%%%%%%%%%%%%%%%%
\appendix

\section{Details of the calculations for bi-directional mutation}
\label{sec:App_mut}

Staring from Eq. (\ref{revmut:2}) in Sec. \ref {subsec:mut}, we introduce the probability generating function (p.g.f.) $ G(z_1,z_2,t) = \sum_{p,q = 0}^{\infty} z_1^p z_2^q P_{p,q}(t)$, and find its {\it linear} partial differential equation (PDE) as below:

\begin{eqnarray}
\frac{\partial G}{\partial t} &=& \frac{\partial G}{\partial z_1}[\beta_1 z_1(z_1-1) +\mu_1 z_1(z_2-1)]+ \frac{\partial G}{\partial z_2}[\beta_2 z_2(z_2-1) +\mu_2 z_2(z_1-1)]\nonumber \\
 \label{revmut:3}
\end{eqnarray}

Next  the cumulant generating function is used, which is defined as $K(\theta_1,\theta_2,t) = \ln [G(z_1 = e^{\theta_1}, z_2 = e^{\theta_2}, t)] = \langle e^{\theta_1 p + \theta_2 q} \rangle$. Using this definition we can transform the PDE for $G(z_1,z_2,t)$ into a PDE for $K(\theta_1,\theta_2,t)$ and then substitute an expansion for $K$, wherein 
\begin{equation}
 K =  \sum_{i+j \geq 1}^{\infty} [(\theta_1^i \theta_2^j)/(i! j!)] \kappa_{i,j}(t).
\end{equation}
 Here $\kappa_{i,j}(t)$ are the cumulants of the joint probability distribution. In particular the first few cumulants are given by  $\kappa_{1,0} = \langle n \rangle$, $\kappa_{0,1} = \langle x \rangle$, $\kappa_{2,0} = Var[n]$, $\kappa_{0,2} = Var[x]$ and $\kappa_{1,1} =\langle n x \rangle - \langle n\rangle \langle x\rangle $ . Equating the coefficients of $\theta_1,\theta_2,\theta_1\theta_2, \theta_1^2$ and $\theta_2^2$, we obtain a system of {\it coupled linear}  ordinary differential equations (ODEs) for the cumulants as below:
\begin{eqnarray}
{\kappa^{'}_{1,0}}(t)&=& \beta_1 \kappa_{1,0}+\mu_2 \kappa_{0,1}\nonumber \\
{\kappa^{'}_{0,1}}(t)&=& \beta_2 \kappa_{0,1}+\mu_1 \kappa_{1,0}\nonumber \\
{\kappa^{'}_{1,1}}(t)&=& (\beta_1+\beta_2)\kappa_{1,1}+\mu_1 \kappa_{2,0}+\mu_2\kappa_{0,2} \nonumber\\
{\kappa^{'}_{2,0}}(t)&=&\beta_1 \kappa_{1,0}+2 \beta_1\kappa_{2,0}+2\mu_2\kappa_{1,1}\nonumber\\
{\kappa^{'}_{0,2}}(t)&=&(\beta_2+\mu_2)\kappa_{0,1}+2 \beta_2\kappa_{0,2}+\mu_1\kappa_{1,0}+2\mu_1\kappa_{1,1}
\label{revmut:4}
\end{eqnarray}
In above equation, primes denote the ordinary time-derivatives. Note that the linear PDE of p.g.f. $G$ ultimately leads to  linear cumulant equations, which is a property of a linear process like the process considered here. Moreover, by putting $\mu_2 =0$ in Eqs. (\ref{revmut:3}) and (\ref{revmut:4}), we get back the corresponding equations in the original Bertlett's formulation of the LD model  \citep[see][pg. 23] {qizheng}. Solving the system of ODEs given by Eq. (\ref{revmut:4}) we obtain the means and variances of the populations  as follows:
\begin{eqnarray}
\langle n \rangle &=& A_1 e^{\alpha_1 t} + A_2 e^{\alpha_2 t}
\nonumber \\ \langle x \rangle &=& B_1 e^{\alpha_1 t} + B_2
e^{\alpha_2 t} \nonumber \\ Var[n] &=& C_1 e^{2 \alpha_1 t} + C_2
e^{2\alpha_2 t} + C_3 e^{\alpha_1 t} + C_4 e^{\alpha_2 t}+C_5
e^{(\beta_1+\beta_2) t} \nonumber\\ Var[x] &=& C_1^{'} e^{2 \alpha_1
  t} + C_2^{'} e^{2\alpha_2 t} + C_3^{'} e^{\alpha_1 t} + C_4^{'}
e^{\alpha_2 t}+C_5^{'} e^{(\beta_1+\beta_2)
  t}\nonumber\\ \text{where,}~\alpha_1 &=& [ (\beta_1 + \beta_2) +
  \sqrt{(\beta_1 - \beta_2)^2 + 4 \mu_1 \mu_2}]/2,\nonumber\\ \alpha_2
&=& [ (\beta_1 + \beta_2) - \sqrt{(\beta_1 - \beta_2)^2 + 4 \mu_1
    \mu_2}]/2.
\label{revmut:moments}
\end{eqnarray}
Here $A_1,\cdots,B_2$ and $C_1,\cdots,C_5^{'}$ are constants
involving the parameters $\beta_1, \beta_2, \mu_1, \mu_2, n(0)$ and
$x(0)$. Since in Eq. (\ref{revmut:moments}), $\alpha_1$ is always greater
 than $\alpha_2$, we arrive at the asymptotic expressions of the means in Eq. (\ref{revmut:mean}) of Sec. \ref{subsec:mut}. Again, noting that the rate $2 \alpha_1$ dominates over the other rates $\alpha_1,\alpha_2$, $2\alpha_2$ and $(\beta_1+\beta_2)$, appearing in Eq. (\ref{revmut:moments}), we get the asymptotic results of the variances in Eq. (\ref{revmut:var}) of Sec. \ref{subsec:mut}.

%%%%%%%%%%%%%%%%%%%%%%%%%%%%%%%%%%%%%%%%%%%%%%%%%%%%%%%%%%%%%%%%%%%%%%%%%%%

\section{Details of the calculations for the Lysis-Lysogeny model}
\label{sec:app_lysis}
For the Lysis-Lysogeny model defined in Eq. (\ref{lysis:1}), the non-linear Master equation is:
\begin{eqnarray}
\frac{\partial P_{n,x,v}(t)}{\partial t} &=& \beta_1 (n-1)P_{n-1,x,v}(t) + \beta_2(x-1) P_{n,x-1,v}(t) \nonumber\\
&&+\phi f (n+1)(v-\alpha+1) P_{n+1,x,v-(\alpha-1)}(t) \nonumber\\
&&+\phi (1-f)(n+1)(v+1)  P_{n+1,x-1,v+1}(t)\nonumber\\
&&-(\beta_1 n +\beta_2 x +\phi n v)P_{n,x,v}(t)
\label{appB:1}
\end{eqnarray}
Under the assumption of $v=constant$ the ecological model of Eq. (\ref{lysis:1}) exactly reduces to the model of  Eq. (\ref{lysis:2}) and the corresponding Master equation becomes linear as follows:
\begin{eqnarray}
\frac{\partial P_{n,x}(t)}{\partial t} &=& \beta_1 (n-1)P_{n-1,x}(t) + \beta_2(x-1) P_{n,x-1}(t) +\phi^{'}f (n+1)  P_{n+1,x}(t) \nonumber\\
&&+\phi^{'}(1-f)(n+1)  P_{n+1,x-1}(t)-(\beta_1 n +\beta_2 x +\phi^{'} n)P_{n,x}(t)
\label{appB:2},
\end{eqnarray}
where $\phi^{'}=\phi v=constant$. Following similar steps as in Appendix A, from the above equation we get the ODEs for the cumulants as below:
\begin{eqnarray}
{\kappa^{'}_{1,0}}(t)&=& \beta_1^{'} \kappa_{1,0} \nonumber \\
{\kappa^{'}_{0,1}}(t)&=& \beta_2 \kappa_{0,1}+\mu^{'} \kappa_{1,0}\nonumber \\
{\kappa^{'}_{1,1}}(t)&=& (\beta_1^{'}+\beta_2)\kappa_{1,1}+\mu^{'} \kappa_{2,0}-\mu^{'} \kappa_{1,0} \nonumber\\
{\kappa^{'}_{2,0}}(t)&=& (\beta_1+\phi^{'}) \kappa_{1,0}+2 \beta_1^{'} \kappa_{2,0}\nonumber\\
{\kappa^{'}_{0,2}}(t)&=&\beta_2 \kappa_{0,1}+2 \beta_2\kappa_{0,2}+\mu^{'} \kappa_{1,0}+2\mu^{'} \kappa_{1,1}
\label{appB:3}
\end{eqnarray}
Here $\beta_1^{'} \equiv (\beta_1 - \phi^{'})$ and $\mu^{'} \equiv  \phi^{'}(1-f)$. We recognize that $\beta_1^{'}$ is  analogous to the birth rate of the wildtype,  while $\mu^{'}$ is analogous to the mutation rate in the original LD model. The explicit results for the moments are:
\begin{eqnarray}
\langle n \rangle &=&  n(0) e^{\beta_1^{'} t}  \nonumber \\
\langle x \rangle &=& x(0) e^{\beta_2 t} + [\mu^{'}n(0)/(\beta_1^{'}-\beta_2)] [e^{\beta_1^{'} t} - e^{\beta_2 t}] \nonumber \\
Var[n] &=& [(\beta_1+\phi^{'})n(0)/\beta_1^{'}] [e^{2 \beta_1^{'} t} - e^{\beta_1^{'} t}]\nonumber\\
Var[x] &=& D_1 e^{2 \beta_1^{'} t}+D_2 e^{2\beta_2 t}+D_3 e^{\beta_1^{'} t}+D_4 e^{\beta_2 t}+D_5 e^{(\beta_1^{'}+\beta_2) t}
\label{appB:4}
\end{eqnarray}
where the coefficients $D_1,...,D_5$ are constants involving the parameters $\beta_1, \beta_2, \mu^{'}, \phi^{'}, n(0)$ and $x(0)$.

%%%%%%%%%%%%%%%%%%%%%%%%%%%%%%%%%%%%%%%%%%%%%%%%%%%%%%%%%%%%%%%%%%%%%%%%%%%%%%

\section{A proposed experimental protocol to map a lysogeny process to a mutation process}
\label{sec:protocol}

Our motivation for the reduction of a Lysis-Lysogeny model to a mutation-like ecological model comes from the experimental
procedure of \citet{kourilsky73}. In this experiment, the reinfection of
cells by phages produced in lysis was prevented by using two different
broths. At first, sensitive cells were allowed to grow in {\it
  tryptone maltose} broth, where most of the infection events took
place.  Then, these infected cells were transferred into {\it Hershey
  citrate} broth where phage-adsorption was severely reduced, and the
phages were diluted out. This leads us to suppose that the removal of
unabsorbed free phages by repeatedly diluting the cultures may be
possible experimentally, and this may keep the $V$ population more or
less constant. Based on this, we suggest that if one can periodically remove the free phages after equal time intervals $T$, then  the viral population $v(t)$ may become quasi-stationary. We show in Sec. \ref{sec:lysis} that it works at least numerically.

The curious reader may note that the length of the time-window over which the LD like behavior is noticeable depends on the choice of $T$. The time window terminates by an avalanche in viral number and a sharp decrease of the population of the sensitive cells which eventually go to extinction. The reason behind this phenomenon can be summarized as follows. 

In kinetic Monte-Carlo
the average time gap $\langle \tau \rangle$ between any two successive
reactions described in Eq. (\ref{lysis:1}) should be much greater than
the chosen time $T$. Otherwise too many reactions will occur within
$T$ which can violate the condition $v(t)\approx constant$. But as
$\langle \tau \rangle = 1/((\beta_1 + \phi^{'})n +\beta_2 x)$ is
actually a decreasing function of time, at some point $T >> \langle
\tau \rangle$ and frequent lytic bursts guarantee a catastrophic
proliferation in viral number. Thus a choice of small enough $T$ is
important.

%%%%%%%%%%%%%%%%%%%%%%%%%%%%%%%%%%%%%%%%%%%%%%%%%%%%%%%%%%%%%%%%%%%%%%%%%%%%%%

\section{Description of the model of four-species HGT}
\label{sec:App_HGT}

The original model of \citet{Goldenfeld08} was treated 
 at a mean-field (deterministic) level, and 
birth of species were ignored.  In contrast, here we allow unrestricted
birth of each species in the presence of unlimited resources, and
treat the four-species HGT as a stochastic process.  
In this model  the genomes have an active locus
gene and another storage locus gene, either of which can be $0$ or
$1$.  Thus four possible genotype populations can arise:
$N_{00}$, $N_{01}$, $N_{10}$ and $N_{11}$, where the first index in the
subscript denotes the active gene and the second index denotes the
storage gene. Let $n_{00}(t)$, $n_{01}(t)$, $n_{10}(t)$ and
$n_{11}(t)$ denote their population sizes respectively.  These
populations interact with each other via a shared gene pool. The
probability of getting a $0$ gene is
$p_0=(2n_{00}+n_{01}+n_{10})/2n_{tot}$, where
$n_{tot}=n_{00}+n_{01}+n_{10}+n_{11}$. Conversely, the probability of
getting a $1$ gene is $p_1=1-p_0$. After having a gene $i$ ($i=0$ or
$1$) from the shared gene pool, this gene can recombine with the
active locus with probability $q_i$, while it can recombine with the
inactive storage locus with probability $(1- q_i)$. The full
stochastic processes along with the births of the four species can be
described by the following reactions:
\begin {eqnarray}
R_1: &&~~ N_{00} \stackrel{\beta_1}{\rightarrow} N_{00} + N_{00} \nonumber \\
R_2: &&~~  N_{01}  \stackrel{\beta_2}{\rightarrow}  N_{01} +  N_{01} \nonumber \\
R_3: &&~~  N_{10}  \stackrel{\beta_3}{\rightarrow}  N_{10} +  N_{10} \nonumber \\
R_4: &&~~  N_{11}  \stackrel{\beta_4}{\rightarrow}  N_{11} +  N_{11} \nonumber\\ 
R_5: &&~~ N_{00} \stackrel{r p_1(1-q_1)}{\xrightarrow{\hspace*{0.8cm}}}N_{01} \nonumber \\
R_6: &&~~ N_{00} \stackrel{r p_1q_1}{\xrightarrow{\hspace*{0.8cm}}}N_{10} \nonumber \\
R_7: &&~~ N_{01} \stackrel{r p_0(1-q_0)}{\xrightarrow{\hspace*{0.8cm}}}N_{00} \nonumber \\
R_8: &&~~ N_{01} \stackrel{r p_1q_1}{\xrightarrow{\hspace*{0.8cm}}}N_{11} \nonumber \\
R_9: &&~~ N_{10} \stackrel{r p_0q_0}{\xrightarrow{\hspace*{0.8cm}}}N_{00} \nonumber \\ 
R_{10}: &&~~ N_{10} \stackrel{r p_1(1-q_1)}{\xrightarrow{\hspace*{0.8cm}}}N_{11} \nonumber \\
R_{11}: &&~~ N_{11} \stackrel{r p_0q_0}{\xrightarrow{\hspace*{0.8cm}}}N_{01} \nonumber \\
R_{12}: &&~~ N_{11} \stackrel{r p_0(1-q_0)}{\xrightarrow{\hspace*{0.8cm}}}N_{10} 
\label{Goldenfeld:1}
\end{eqnarray}

\begin{figure*}
\centering
 \includegraphics[width=0.7\textwidth]{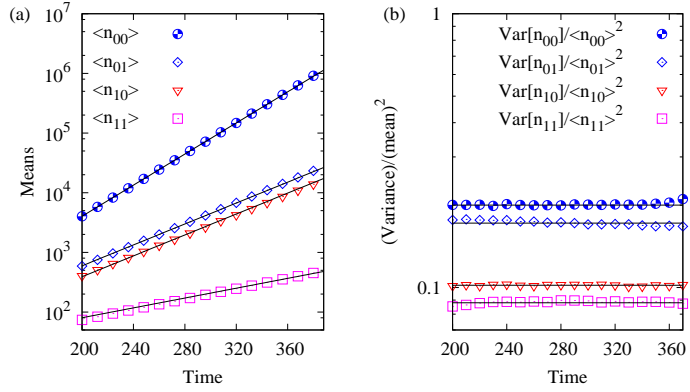}
\caption{(a) Linear-log plot of $\langle n_{00} \rangle$, $\langle
  n_{01} \rangle$, $\langle n_{10} \rangle$ and $\langle n_{11}
  \rangle$ versus time $t$ for four-species HGT. Data from kinetic
  Monte-Carlo simulation are shown in various symbols as
  indicated. The fitted exponential functions (solid continuous
  straight lines) are: $\langle n_{00} \rangle \sim 10.0 \times
  \exp(\lambda_1 t)$; $\langle n_{01}\rangle \sim 10.5 \times \exp(\lambda_2
  t)$; $\langle n_{10}\rangle \sim 8.6 \times \exp(\lambda_2 t)$ and
  $\langle n_{11}\rangle \sim 12.0 \times \exp(\lambda_4 t)$ (see 
  Eq. (\ref{Goldenfeld:3}) in Appendix C). (b)
  Linear-log plot of $Var[n_{00}]/\langle n_{00} \rangle^2$,
  $Var[n_{01}]/\langle n_{01} \rangle^2$, $Var[n_{10}]/\langle n_{10}
  \rangle^2$ and $Var[n_{11}]/\langle n_{11} \rangle^2$ versus time
  $t$ for four-species HGT. The straight lines show that the data are constant 
   with time. The  data are for $n_{00}(0)=n_{01}(0)=
  n_{10}(0)=n_{11}(0)= 10$, $\beta_1=0.03$, $\beta_2=\beta_3=0.02$,
  $\beta_4=0.01$, $r=0.0005$, $q_0=0.9$ and $q_1=0.1$. The data are obtained by averaging over $10^5$ independent histories. }
\label{fig:4}  % Give a unique label
\end{figure*}
%%%%%%%%%%%%%%%%%%%%%%%%%%%%%%%%%%%%%%%%%%%%%%%%%%%%%%%%%%%%

The reactions $R_1$ to $R_4$ are linear birth processes and the
reactions $R_5$ to $R_{12}$ represent the nonlinear interactions
through HGT.  The rate of HGT in this model is $r$ and we will also
assume that $\{\beta_1,\beta_2,\beta_3,\beta_4 \}>>r$. The Master
equation is non-linear (in population sizes) in general. However at large times, the
nonlinear terms may be approximated to linear ones for the choice of
parameters: $n_{00}>>\{n_{01},n_{10},n_{11}\}$, $n_{00}>>n_{01}^2$,
$\beta_1>\{ \beta_2,\beta_3,\beta_4\}$, and $0<q_i<1/2$,
$1/2<q_{1-i}<0$ (where $i=0$ or $1$). From the approximate linear
Master equation we get the ODEs for the moments and solving them 
get the following solutions for the means:
\begin{eqnarray}
\langle n_{00} \rangle &\approx& c_1 e^{\lambda_1t}+c_2 e^{\lambda_2t}+c_3 e^{\lambda_3t}+c_4 e^{\lambda_4t} \nonumber\\
\langle n_{01} \rangle &\approx& k_1 e^{\lambda_2t}+k_2 e^{\lambda_3t}+k_3 e^{\lambda_4t} \nonumber\\
\langle n_{10} \rangle &\approx& k_1^{'} e^{\lambda_2t}+k_2^{'} e^{\lambda_3t}+k_3^{'} e^{\lambda_4t} \nonumber\\
\langle n_{11} \rangle &\approx& k e^{\lambda_4 t}, \nonumber\\
&&\text{where}\nonumber\\
\lambda_1 &=& \beta_1 ~~,~~\lambda_2 = \left(\frac{\beta_2+\beta_3}{2} - \frac{r}{4} \right) +\sqrt{b}\nonumber\\
\lambda_3 &=& \left(\frac{\beta_2+\beta_3}{2} - \frac{r}{4} \right) -\sqrt{b} ~~,~~\lambda_4 = \beta_4 -r \nonumber\\
b&=& r^2+4(\beta_2-\beta_3-r+2r q_0) (\beta_2-\beta_3+2r(q_0-q_1)) 
\label{Goldenfeld:3}
\end{eqnarray}
Here $c_1,\cdots,k$ are constants. Note that $\lambda_2>\lambda_3$
always, but nothing can be concluded about the relative magnitudes of
$\lambda_2$ and $\lambda_4$, which will depend on the exact parameter
values. Thus two cases can arise at large times: (i) for $\lambda_2 >
\lambda_4$, we have $\langle n_{00} \rangle \sim e^{ \lambda_1t}$,
$\langle n_{01} \rangle \sim \langle n_{10}\rangle \sim e^{\lambda_2
  t}$ and $\langle n_{11} \rangle \sim e^{ \lambda_4t}$; while (ii)
for $\lambda_2 < \lambda_4$, we get $\langle n_{00}\rangle \sim e^{
  \lambda_1t}$ and $\langle n_{01}\rangle \sim \langle n_{10}\rangle
\sim \langle n_{11} \rangle \sim e^{ \lambda_4t}$.

 We exactly simulate the
reactions of Eq. (\ref{Goldenfeld:1}) using kinetic Monte-Carlo. In certain regions of parameter space, at large times,
we do find LD like behavior with the associated GNF property.  We show
the data for means and variances at large times in Fig. 5,
for one such parameter and number regime defined by the following
relations: (i) $\beta_1>\{ \beta_2,\beta_3,\beta_4\}$, (ii)
$0<q_i<1/2$ and $1/2<q_{1-i}<0$, where $i=0$ or $1$, (iii)
$n_{00}>>\{n_{01},n_{10},n_{11}\}$, and (iv) $n_{00}>>n_{01}^2$. The GNF property holds for each species as is clearly evident from Fig. 5b. We also find a
match between our data and the approximate analytical answer (Fig. 5a).
We note that unlike the two-species HGT, for the four-species HGT the
reduction to LD like behavior may not be true in general for all
regions of the parameter space.

Similar analysis as above holds for two other parameter and number regimes : (1) $\beta_4>\{ \beta_1,\beta_2,\beta_3\}$, $n_{11}>>\{n_{01},n_{10},n_{00}\}$ and $n_{00}>>n_{01}^2$; (2) $n_{11}\sim n_{01}\sim n_{10}\sim n_{00}$.

\bibliographystyle{model2-names}
\bibliography{draft_5th_sub}

%\bibliographystyle{Harvard/agsm}

%% not want to use model2-names.bst.



%% \bibitem must have one of the following forms:
%%   \bibitem[Jones et al.(1990)]{key}...
%%   \bibitem[Jones et al.(1990)Jones, Baker, and Williams]{key}...
%%   \bibitem[Jones et al., 1990]{key}...
%%   \bibitem[\protect\citeauthoryear{Jones, Baker, and Williams}{Jones
%%       et al.}{1990}]{key}...
%%   \bibitem[\protect\citeauthoryear{Jones et al.}{1990}]{key}...
%%   \bibitem[\protect\astroncite{Jones et al.}{1990}]{key}...
%%   \bibitem[\protect\citename{Jones et al., }1990]{key}...
%%   \harvarditem[Jones et al.]{Jones, Baker, and Williams}{1990}{key}...
%%


% \end{thebibliography}

%%%%%%%%%%%%%%%%%%%%%%%%%%%%%%%%%%%%%%%%%%%%%%%%%%%%%%%%%%%%%%%%%%%%%%%%%%%%%%

%%%%%%%%%%%%%%%%%%%%%%%%%%%%%%%%%%%%%%%%%%%%%%%%%%%%%%%%%%%%

\end{document}